\documentclass[aps,prl,twocolumn,groupedaddress]{revtex4}
\usepackage{amssymb}
\usepackage{graphicx}

\begin{document}

\title{Electronic structure and exotic exchange splitting in spin-density-wave states of BaFe$_2$As$_2$}

\author{L. X. Yang$^{1}$, Y. Zhang$^{1}$, H. W. Ou$^{1}$,  J. F. Zhao$^{1}$, D. W. Shen$^{1}$, B. Zhou$^{1}$, J. Wei$^{1}$,
F. Chen$^{1}$, M. Xu$^{1}$, C. He$^{1}$, Y. Chen$^{1}$, Z. D.
Wang$^{1,2}$, X. F. Wang$^3$, T. Wu$^3$, G. Wu$^3$,  X. H. Chen$^3$,
M. Arita$^{4}$, K. Shimada$^{4}$, M. Taniguchi$^{4}$, Z. Y.
Lu$^{5}$, T. Xiang$^{6}$ and D. L. Feng$^{1}$}
\email{dlfeng@fudan.edu.cn}

\affiliation{$^1$Department of Physics, Surface Physics Laboratory
(National Key Laboratory), and Advanced Materials Laboratory, Fudan
University, Shanghai 200433, P. R. China}

\affiliation{$^2$Department of Physics, The University of Hong Kong,
Hong Kong, P. R.  China.}

\affiliation{$^3$Hefei National Laboratory for Physical Sciences at
Microscale and Department of Physics, University of Science and
Technology of China, Hefei, Anhui 230026, P. R. China}

\affiliation{$^4$Hiroshima Synchrotron Radiation Center and Graduate
School of Science, Hiroshima University, Hiroshima 739-8526, Japan.}

\affiliation{$^5$Department of Physics, Renmin University of China,
Beijing 100872, P. R.  China. }

\affiliation{$^6$Institute of Physics, Chinese Academy of Sciences,
Beijing 100190, P. R.  China.}

\date{\today}

\begin{abstract}

The magnetic properties in the parent compounds are often intimately
related to the microscopic mechanism of superconductivity. Here we
report the first direct measurements on the electronic structure of
a parent compound of the newly discovered iron-based superconductor,
BaFe$_2$As$_2$, which provides a foundation for further studies. We
show that the energy of the spin density wave (SDW) in
BaFe$_2$As$_2$ is lowered through exotic exchange splitting of the
band structure, rather than Fermi surface nesting of itinerant
electrons. This clearly demonstrates that a metallic SDW state could
be solely induced by interactions of local magnetic moments,
resembling the nature of antiferromagnetic order in cuprate parent
compounds.

\end{abstract}

\pacs{74.25.Jb,74.70.-b,79.60.-i,71.20.-b}

\maketitle


The discovery of superconductivity in iron-pnictide has generated
another intensive wave of research on high temperature
superconductivity \cite{JACS,ChenNature,NLWang1,ZXZhao1,ZXZhao2}.
The record superconducting transition temperature ($T_c$) has been
quickly raised to 56K in LnO$_{1-x}$F$_y$FeAs (Ln=La, Sm, Nd, etc.),
and a $T_c$ of 38K has been reported in Ba$_{1-x}$K$_x$Fe$_2$As$_2$
\cite{bilayer0,Chenbilayer1}. Intriguingly, like in the cuprates,
the ground state of the parent compound LaOFeAs is a magnetically
ordered state, where a spin density wave (SDW) emerges following a
structural transition \cite{PCDai}. Similarly,  BaFe$_2$As$_2$
enters the SDW phase at the transition temperature $T_S$ of around
138K \cite{Chenbilayersingle,WBao1}. Currently, it is unclear
whether the SDW facilitates the electron pairing as the
antiferromagnetic fluctuations arguably do in cuprates, or acts as a
competing order as the charge density wave does in transition metal
dichalcogenides. Therefore, it is crucial to reveal the nature of
the SDW and its manifestation on the electronic structure.

We here report angle resolved photoemission spectroscopy (ARPES)
data of BaFe$_2$As$_2$ single crystals. In the paramagnetic state,
we found that the Fermi surfaces are consisted of two hole pockets
around the Brillouin zone center $\Gamma$ of the tetragonal unit
cell, and one electron pockets at the zone corner M. This
qualitatively resembles the density functional theory (DFT)
electronic structure calculations on LnOFeAs
\cite{LDA0,LDA1,LDA2,LDA3,LDA4} and BaFe$_2$As$_2$ \cite {LDA5}. In
addition, we observed renormalization of the band structure due to
the correlation effects. Below $T_S$, band splitting and folding are
observed, which induces several additional Fermi surfaces. We do not
observe any energy gap in the SDW state, therefore, the SDW in
iron-pnictides is unlikely driven by the Fermi surface nesting that
is now commonly believed to account for SDW in metals like Chromium
\cite{chromium}. On the other hand, we show that the exchange
splitting caused by the effective local moments is responsible for
the observed band splitting, and it does lower the total electronic
energy in the SDW state. It suggests that the magnetic order in the
metallic parent compound of iron-pnictides partially shares the
local nature of various magnetic orders in Mott insulators.

\begin{figure*}[t]
\includegraphics[width=16cm]{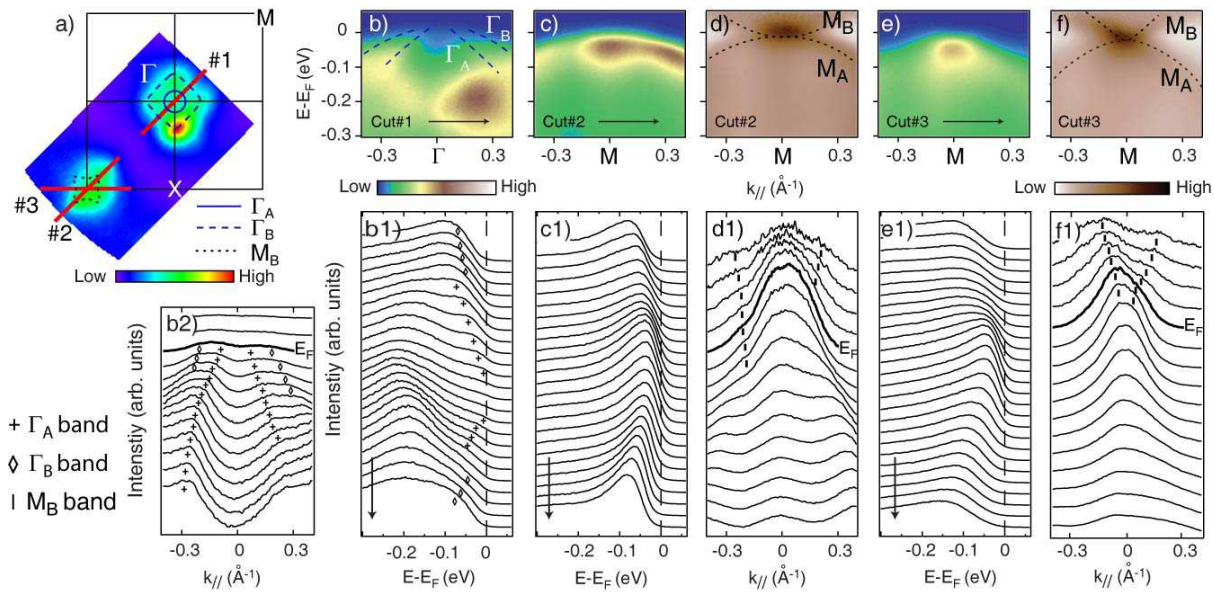}
\caption{(color) Normal state Fermi surface and band dispersion of
BaFe$_2$As$_2$. (a) Photoemission intensity map in the Brillouin
zone at $E_F$. The dashed lines are the measured Fermi surfaces. (b)
Photoemission intensity along cut \#1 through $\Gamma$ as indicated
in panel a, and the corresponding (b1) EDC's, and (b2) MDC's. (c)
Photoemission intensity along cut \#2 through the zone corner M, and
(c1) the corresponding EDC's. (d) shows the same data as in panel c
but after the individual MDC's are normalized by its total weight,
and (d1) shows some of the normalized MDC's in (d). (e)
Photoemission intensity along cut \#3 through the zone corner M, and
(e1) the corresponding EDC's. (f) shows the same data as in panel e
but after the individual MDC's are normalized by its total weight,
and (f1) shows some of the normalized MDC's. The energy difference
between two neighboring MDC's is 9meV. Data were taken at 160K with
21.2 eV photons at the synchrotron.}
\end{figure*}

The BaFe$_2$As$_2$ single crystals were synthesized and
characterized as described in ref.9 with $T_S=138K$. ARPES
measurements were performed with photons from beamline 9 of
Hiroshima synchrotron radiation center, and 21.2eV photons from a
helium-discharge lamp. Scienta R4000 electron analyzers are equipped
in both setups. The overall energy resolution is 10meV, and angular
resolution is 0.3 degree. The samples were cleaved \textit{in situ},
and measured under ultra-high-vacuum of
$3\times10^{-11}$\textit{torr} at the synchrotron, and
$1\times10^{-10}$\textit{torr} in the helium lamp system. The sample
surface quality is confirmed by low energy electron diffraction.

\begin{figure*}[t]
\includegraphics[width=15cm]{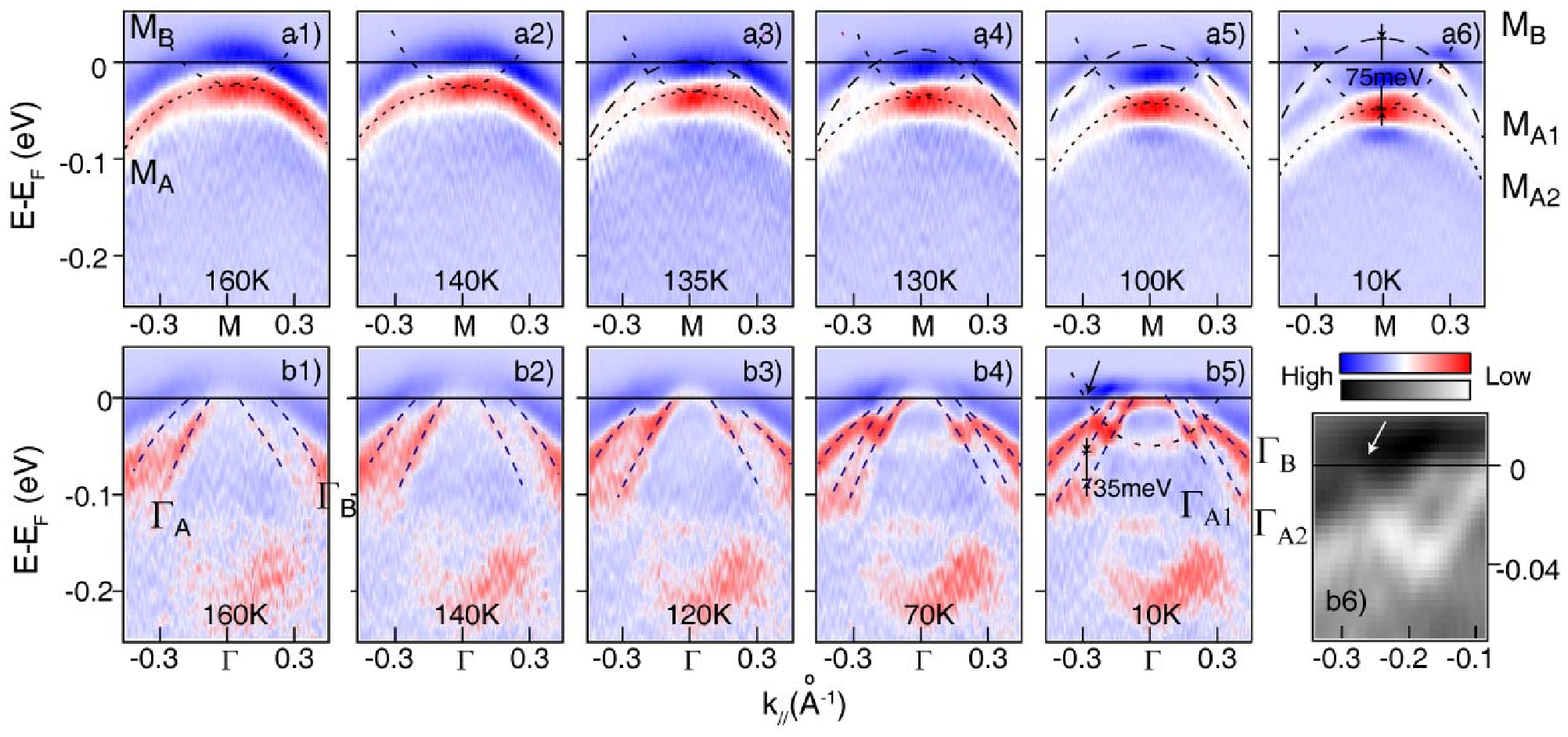}
\caption{(color) Temperature dependence of the electronic structure
of BaFe$_2$As$_2$. Second derivative of photoemission intensity with
respect to energy (a1-a6) along cut \#2 in Fig.1a at 160K, 140K,
135K, 130K, 100K, and 10K respectively, and (b1-b5) along cut \#1 in
Fig.1a at 160K, 140K, 120K, 70K, and 10K respectively. (b6) is an
enlarged view of data in b5 near the Fermi crossing. Data were taken
at the synchrotron with 21.2eV photons. Dashed lines are the guides
of eye for the bands. Note the minimum of the second derivative
represents a peak, thus the lower part (red or white color)
represents the band. }
\end{figure*}

Photoemission intensity at the Fermi energy ($E_F$) is a direct
measure of the Fermi surfaces. Fig.1a shows the normal state map of
the photoemission intensity: there are features around both the
$\Gamma$ and M points. Figs.1b, 1b1, 1b2 illustrate the
photoemission intensity, energy distribution curves (EDC's) and
momentum distribution curves (MDC's) across the momentum cut \#1 in
Fig.1a through $\Gamma$. Two bands (assigned as $\Gamma_A$ and
$\Gamma_B$ band respectively) could be identified to cross $E_F$,
giving two hole-type Fermi surfaces around $\Gamma$, as predicted by
the band structure calculations
\cite{LDA0,LDA1,LDA2,LDA3,LDA4,LDA5,NLWang2}. In addition, there is
a feature near 200meV below $E_F$. On the other hand, Figs.1c and
1c1 illustrate the photoemission intensity and EDC's along the cut
\#2 in Fig.1a through M. The spectra are dominated by a strong
feature dispersing toward M. However,  a weak feature near $E_F$
could be resolved in the normalized MDC's (Fig.1d1, see caption for
description). Since the peak positions of the MDC's are not altered,
one could determine the dispersion by tracking the features on the
normalized image (Fig.1d). We found two bands in Fig. 1d (see dashed
lines): the band at higher binding energies is below $E_F$ (named as
$M_A$ band), while the other one crosses $E_F$ (named as $M_B$ band,
also shown by marks in Fig. 1d1), which corresponds to the
electron-like pocket around M in Fig.1a. Figs.1e-f1 illustrate
similar behavior along the MX direction (cut \#3), except that the
Fermi crossings are closer.

The SDW effects are examined by the temperature dependence of the
bands, which could be tracked in the second derivative of the
photoemission intensity with respect to energy (Fig.2). Above $T_S$,
Figs.2a1 and 2a2 illustrate mainly a feature for the $M_A$ band, and
the $M_B$ band is very weak, as depicted in Fig.1. Once entering the
SDW state, we clearly observed the splitting of the $M_A$ band into
the upper $M_{A1}$ band, and the lower $M_{A2}$ band, as shown by
the dashed lines in Fig.2a3. With decreased temperature, this
splitting increases quickly, and eventually reaches about 75meV.
Moreover, the $M_B$ band also continuously goes deeper in binding
energy. This dramatic temperature dependence and the sharp
correlation with $T_S$ prove that the electronic structure measured
here reflects the bulk properties of the system with the same doping
concentration.

The temperature evolution of the bands around $\Gamma$ is shown in
Fig.2b1-b5. Below $T_S$, $\Gamma_A$ band splits into $\Gamma_{A1}$
(closer to $\Gamma$) and $\Gamma_{A2}$ bands. The Fermi crossing of
the $\Gamma_{A2}$ band is pushed away from $\Gamma$ with decreasing
temperature, while $\Gamma_{A1}$ is slightly pushed downward in
energy. The maximum splitting is about 35 meV in Fig.2b5. Meanwhile,
the top of $\Gamma_B$ band is pushed downward slightly. However,
since the folded $M_B$ bands could be observed around $\Gamma$ as
indicated by the dashed line in Fig.2b5, the shift of the $\Gamma_B$
band is likely due to the hybridization with the folded $M_B$ band.
The original Fermi crossing of $\Gamma_B$ band is undistinguishable
from that of $\Gamma_{A2}$ band. Furthermore, we note that around M
point, one cannot identify whether the $M_B$ band is gapped or not;
but as shown in Figs.2b5-2b6, the Fermi crossing of the folded $M_B$
band around $\Gamma$(highlighted by an arrow) decisively rules out
any gap opening. This is a crucial observation, because the presence
of gap on Fermi surface is a natural result of the so-called
``Fermi-surface-nesting" mechanism, which is prototypical for SDW in
itinerant electron systems like Chromium and its alloys
\cite{chromium}. Since the Fermi surfaces around M and $\Gamma$ are
roughly apart by the SDW ordering wavevector $(\pi, \pi)$, the
nesting between them was suggested to lead to the SDW formation
\cite{NLWang2}. Therefore, the \textit{absence} of gaps for all the
bands declares likely inapplicability of the Fermi surface nesting
as a mechanism for SDW in iron pnictides.

\begin{figure}[t]
\includegraphics[width=8.5cm]{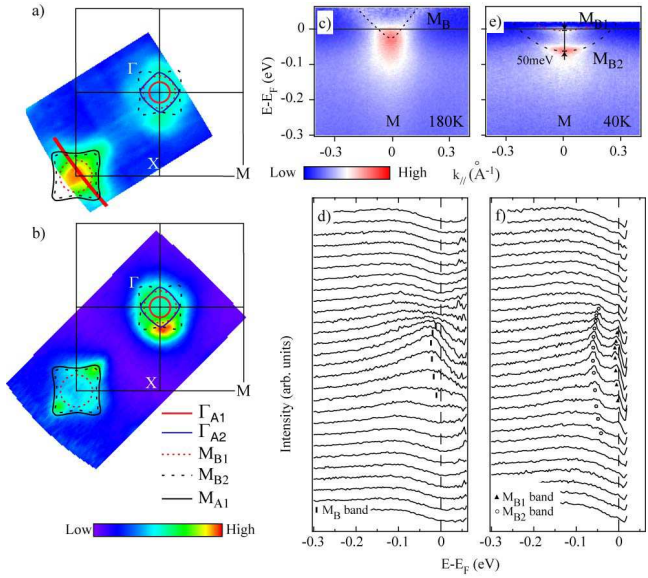}
\caption{(color online) Identification of the splitting of the $M_B$
band of BaFe$_2$As$_2$. Photoemission intensity map taken (a) at 40K
with randomly polarized 21.2eV light from a helium lamp, and (b) at
10K with circularly polarized 21.2eV synchrotron radiation. The
solid and dashed lines are guides of the eye for the measured
hole-type Fermi surfaces and electron-type Fermi surfaces
respectively in the SDW state. (c) Photoemission intensity taken at
180K along the cut in panel a with helium lamp, and (d) the
corresponding EDC's. (e) and (f) present data taken at the same
condition but at 40K. Data in panels c-f have been divided by the
resolution convoluted Fermi-Dirac distribution function to highlight
the dispersions above $E_F$.}
\end{figure}

The upward shift of the $M_{A1}$ band is accompanied by the downward
shift of the $M_{A2}$ band. Therefore, the downward movement of the
$M_B$ band might suggest another band being pushed up. In fact, this
is observed in data taken with a helium lamp, where a strong feature
is present at M in the SDW state (Fig.3a), while it is absent in
Fig.3b for the synchrotron data. In Figs.3c-d, the normal state band
structure taken with the helium lamp only has one dispersing
feature, just like the $M_B$ band in Fig.1d. However, the
photoemission intensity of the $M_A$ bands is strongly suppressed
here, thus the $M_B$ band dispersion is more clearly observed. In
the SDW state (Figs.3e-f), one could clearly observe that the $M_B$
band actually splits into two bands: one band at pushed to high
binding energies as observed before in Fig.2a6, and the other one
pushed to almost just at $E_F$ (named hereafter as $M_{B2}$ and
$M_{B1}$ band respectively). We observed the folding of the $M_{B1}$
band to $\Gamma$ as well (not shown). Furthermore, many photon
energies have been exploited to explore the dispersion perpendicular
to the FeAs plane and possible matrix element effects. We had not
observed much dispersion along $k_z$ direction \cite{Rice}, which
suggests that the coupling along the c-direction is weak, being
consistent with the measured large out-of-plane/in-plane resistivity
ratio of 150 \cite{Chenbilayersingle}. As a result of these
comprehensive studies, the Fermi surfaces in the SDW state of
BaFe$_2$As$_2$ are summarized in Figs.3a-3b, which are much more
complicated than the normal state ones. Around M, there are three
Fermi surfaces: the inner pocket corresponds to $M_{B1}$, and the
Fermi surfaces of $M_{B2}$ and $M_{A1}$ are almost coincident and
barely resolved. Around $\Gamma$, there are four Fermi surfaces
corresponding to the two split $\Gamma_A$ bands, and the folded
$M_{B1}$ and $M_{B2}$ pockets. We note there are probably even more
folded Fermi surfaces, and Fermi surface of the $\Gamma_B$ band that
just could not be resolved at the moment possibly due to various
matrix element effects.

The DFT calculations show that there are only hole-type Fermi
surfaces around $\Gamma$ and an electron-type Fermi surface around M
in the paramagnetic phase. This is consistent with our observations
\cite{LDA5}. However, quantitatively, there are certainly
discrepancies between the data and calculations. For example, the
bottom of the normal state $M_B$ are about 20meV below $E_F$, while
the top of the $\Gamma$ bands is estimated to be at most 50meV above
$E_F$, if the measured dispersion is extrapolated to $\Gamma$. Thus
their separation is about 70meV, while it is approximately
300-400meV in various band calculations, indicating certain
correlation effects that likely missing in the calculations.

The band structure calculation of the collinear SDW state of
BaFe$_2$As$_2$ failed to reproduce the observed band splitting
\cite{LDA5}. The 0.35$\%$ structural distortion of the lattice
constants cannot account for the splitting energy scale either, plus
there are no degenerate bands around M or $\Gamma$ in the
non-magnetic ground state to split with. On the other hand, the
splitting energy scales are comparable with the effective exchange
interactions between the nearest and next-nearest neighbor local
moments, which have been estimated to be 30-35 meV  by comparing
energies of different spin configurations for BaFe$_2$As$_2$
\cite{LDA5}. This strongly suggests the importance of local spin
correlations. In fact, only two known possible causes for such
splitting are left on the table, \textit{i.e.} spin-orbital
coupling, and exchange splitting. The former could be present in the
paramagnetic state like in GaAs, while the latter only exists in the
magnetically ordered state, thus is favored by the observed definite
correlation between the splitting and $T_S$. The exchange splitting
phenomenon has been widely observed in ferromagnets before
\cite{exsplit}. To consider the exchange splitting in an SDW state,
one could split the states into one configuration where spins are in
phase with the SDW order, and the other out of phase with the order.
For simplicity and illustration purpose, Fig.4 views the SDW as two
``ferromagnetic sub-lattices". Taking the spin-up sub-lattice for
example, the bands for majority (up spin) and minority (down spin)
electrons are split in energy due to exchange interactions. Since
the majority band is occupied by more electrons than the minority
band, the total energy would be saved when the energy of the two
spin states are pushed equally in opposite directions. Therefore,
exchange splitting indeed could naturally explain our observations.
Furthermore, it suggests that one could study the problem from the
perspective of effective local moments \cite{LDA6,JPHuPRB}, where
the collinear SDW order is caused by the exchange interactions
between the nearest neighbors and the next nearest neighbors.
Consequently, the SDW is naturally commensurate and does not require
opening of a gap on the Fermi surface, or nesting of the Fermi
surfaces, similar to the antiferromagnetic order in Mott insulators.
This view may receive further support from the recent neutron
scattering study that gives the ordered magnetic moment to be about
0.87$\mu_B$ per Fe ion, which is difficult to be understood in a
pure itinerant spin picture due to small Fermi surfaces. One could
then estimate one essential parameter of magnets, the Stoner ratio
(i.e. the exchange splitting over magnetic moment). It is about
0.1eV/$\mu_B$ for the $M_A$ band of BaFe$_2$As$_2$, which is
anomalously small, being only about 25$\%$ of that of Chromium and
10$\%$ of that of regular ferromagnets like bcc Fe \cite{exsplit}.
This and the band-dependence highlight the exotic SDW properties in
iron-pnictides, demanding further in-depth exploration.

\begin{figure}
\includegraphics[width=8.5cm]{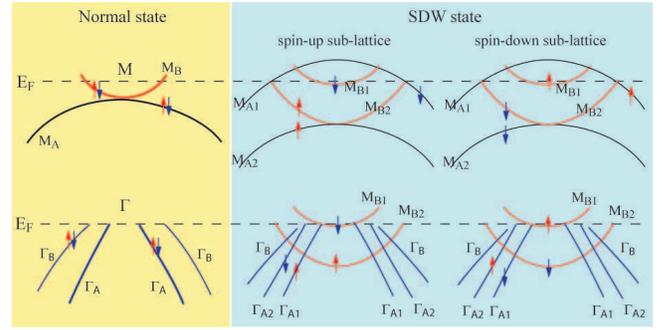}
\caption{(color online) Cartoon of the exchange splitting of the
bands in BaFe$_2$As$_2$.}
\end{figure}%


To summarize, we have measured the electronic structure of an iron
pnictide in detail. Particularly, we have elucidated that there is
no gap at the Fermi surfaces, and SDW is induced by the exchange
interactions between effective local moments. Our results would shed
light on the understanding of the relationship between the SDW and
superconductivity, and set the foundation for further studies in
this field.

We thank Profs. X. F. Jin, J. P. Hu for helpful discussions. This
work was supported by the Nature Science Foundation of China, the
Ministry of Science and Technology of China (National Basic Research
Program No.2006CB921300 and 2006CB922005), and STCSM.


\begin{references}
\bibitem{Rice}At certain photon energies, some more subtle features were observed. Since it
is independent of the arguments made here, we leave it for further
investigations in the future.
\end{references}
\end{document}